\documentclass{mem}
\usepackage{natbib}\usepackage{txfonts}\usepackage{balance}
\usepackage{graphicx}
\usepackage[a4paper]{hyperref}
\idline{75}{282}
\begin{document}
\def\teff{$T\rm_{eff }$} 
\def\kms{$\mathrm {km s}^{-1}$}

\title{
Dynamical evolution of the young stars in the Galactic center
}

   \subtitle{}

\author{
Hagai. B. Perets\inst{1}, Alessia Gualandris\inst{2}, David Merritt\inst{2}
and Tal Alexander\inst{1}}

\institute{
Benoziyo Center for Astrophysics, Weizmann Institute of Science
\and
Center for Computational Relativity and Gravitation, Rochester
Institute of Technology
}

\authorrunning{Perets et al.}

\titlerunning{Galactic center young stars}

\abstract{Recent observations of the Galactic center revealed a nuclear disk
of young OB stars near the massive black hole (MBH), in addition
to many similar outlying stars with higher eccentricities and/or high
inclinations relative to the disk (some of them possibly belonging
to a second disk). In addition, observations show the existence of
young B stars (the 'S-cluster') in an isotropic distribution in the
close vicinity of the MBH ($<0.04$ pc). We use extended N-body simulations
to probe the dynamical evolution of these two populations. We show that the stellar disk could have evolved
to its currently observed state from a thin disk of stars formed in
a gaseous disk, and that the dominant component in its evolution is
the interaction with stars in the cusp around the MBH. We also show
that the currently observed distribution of the S-stars could be 
consistent with a capture origin through 3-body binary-MBH interactions. 
In this scenario the stars are captured at highly eccentric orbits, 
but scattering by stellar black holes could change their eccentricity distribution to be consistent with current observations.
}
\maketitle{}

\section{Introduction}
High resolution observations have revealed the existence of many young
OB stars in the galactic center (GC). Accurate measurements of the
orbital paramters of these stars give strong evidence for the existence
of a massive black hole (MBH) which govern the dynamics in the GC
\citep{sch+02b,ghe+03a}. Most of the young stars are observed in
the central 0.5 pc around the MBH. The young stars population in the
inner 0.04 pc (the 'S-stars' or the 'S-cluster') contain only young
B-stars, in apparently isotropic distribution around the MBH, with
relatively high eccentricities ($0.3\le e\le0.95$) \citep{ghe+03a,eis+05}.
The young stars outside this region contain many O-stars residing
in a stellar disk moving clockwise in the gravitational
potential of the MBH \citep{lev+03,gen+03a,lu+06,pau+06,tan+06}.
The orbits of the stars in this disk have average eccentricity of
$\sim0.35$ and the opening of the disk is $h/R\sim0.1$, where $h$
is the disk height and $R$ is its radius. \citet{pau+06} and \citet{tan+06}
suggested the existence of another stellar disk rotating counter clockwise
and almost perpendicular to the CW disk. This disk
is currently debated as many of the young stars are have intermediate inclinations, and are possibly just outliers that do
not form a coherent disk structure \citep{lu+06}.

Here we briefly report on our study of the dynamical evolution of
the young stars in the GC, both in the stellar disk and in the S-cluster.
We use extensive N-body simulations with realistic number of stars  ($10^{3}-10^{5}$) using the  {\tt gravitySimulator}, a 32-node cluster at the Rochester Institute of Technology that incorporates GRAPE accelerator boards in each of the nodes \citep{har+07}. Thus we are able to probe
the dynamics of the stars near the MBH and their stellar environment.
We study two basic issues: (1) the long term evolution of the S-stars
up to their lifetime of a few $10^{7}$ yrs, including their dynamical
interaction with stars in the vicinity of the MBH; (2) The evolution
of a realistic stellar disk, taking into account both the effects
of non-equal mass stars, as studied earlier, and more importantly
the effect of the interactions of disk stars with the stellar cusp
around the MBH. As we show, the latter component proves to be more
important than the other components discussed in previous studies.
A detailed report of our complete set of out simulations (not shown
here), in addition to analytic calculations will be presented in upcoming
papers (Perets et al., in preparation)

\section{Formation and/or migration origin}

Analytic calculations and simulations have shown that a young stars could have 
formed and grown over short times of thousands to millions of years in a gaseous disk around the MBH (e.g. \citep{nay+05b,lev07}).
Such stars could then form the stellar disk currently observed in
the GC. It was suggested that the 'S-stars' with their very different
properties migrated from the stellar disk through a planetary migration
like process \citep{lev07}. This interesting possibility has not
yet been studied quantitatively, but would suggest that the migrating
stars should have relatively low eccentricities. Another possibility
is that these stars have a different origin, possibly from the disruption
of young binaries and the following capture of one of their components
\citep{gou+03}. It was recently shown that such a scenario could
be consistent with the current knowledge regarding the number of the
observed 'S-stars' \citep{per+06}. The initial eccentricity of the
captured stars should then be very high $(>0.96)$ in this scenario.
We note that other scenarios were suggested for the origin of the
young stars in the GC, but seem to be disfavored by current observations
(see e.g. \cite{ale05} and \cite{pau+06}).

\section{The S-stars}
Only one study has been published of the dynamical evolution of the
      S-stars since their capture / formation, which explored the possible role on IMBH on their evolution \citep{mik+08}.
 Here we present the results of N-body
simulations dedicated to study such evolution. In order to do so we
modeled a small isotropic cusp of 1200 stars, with masses of $3\, M_{\odot}$
(200 stars) and $10\, M_{\odot}$(1000) around a MBH of $3.6\times10^{6}\, M_{\odot}$.
We used a power law radial distribution of $r^{-\alpha}$ extending
from $0.001-0.05$ pc near the MBH, with $\alpha=2$ for the more
massive stars and $\alpha=1.5$ for the lower mass stars. The more
massive stars correspond to the many stellar black holes (SBHs) thought
to exist in this region, whereas the lower mass stars correspond to
the S-stars in the same region. Since some of the S-stars may have
higher masses of $\sim10\, M_{\odot}$, the higher mass stars in the
simulation could also be treated as S-stars. We did not see any major
differences in the evolution of the more massive and the less massive
stars, and we discuss the evolution of both together.

We studied two evolutionary scenarios for the S-stars. In the first
we assumed that the S-stars were captured through a binary disruption
scenario by the MBH \citep{gou+03,per+06} and therefore have initially
highly eccentric orbit $(>0.96)$ and they evolve for few $10^{7}$
yrs. In the second scenario we assumed the S-stars formed in a gaseous
disk and migrated to their current position, and therefore have low
eccentricities ($<0.3$) and they evolved for $5\,$Myrs (the lifetime
of the observed stellar disk. In order to check both scenarios we
followed the evolution of those stars in our simulation with highly
eccentric initial orbits (the first scenario) and those with low eccentricities
(the second scenario) for the appropriate time scales. In Fig. (1)
we show the final eccentricity distribution of the S-stars in both
scenarios, as compared to the the orbits of the observed S-stars (taken
from Gillessen et al.). These results suggest that, given the small
statistics (16 S-stars with known orbits), the first scenario is much
favored since it could explain the currently observed orbits of the
S-stars, i.e. stars on highly eccentric orbits could be scattered
by other stars or SBHs to smaller, and even much smaller eccentricities
during their lifetimes. The second scenario, however, seems to be
excluded (for the given assumptions), since it has major difficulties
in explaining the large number of eccentric orbits in the S-stars
observations vs. the bias towards low eccentricity orbits seen in
the N-body simulations simulations. This is clearly seen both after
$5$ Myrs of evolution, and even after longer evolution, if these
stars formed in an earlier epoch of star formation in a disk (not
currently observed) $20$ Myrs ago.

\begin{figure}[h!]
\resizebox{\hsize}{!}{\includegraphics[clip=true]{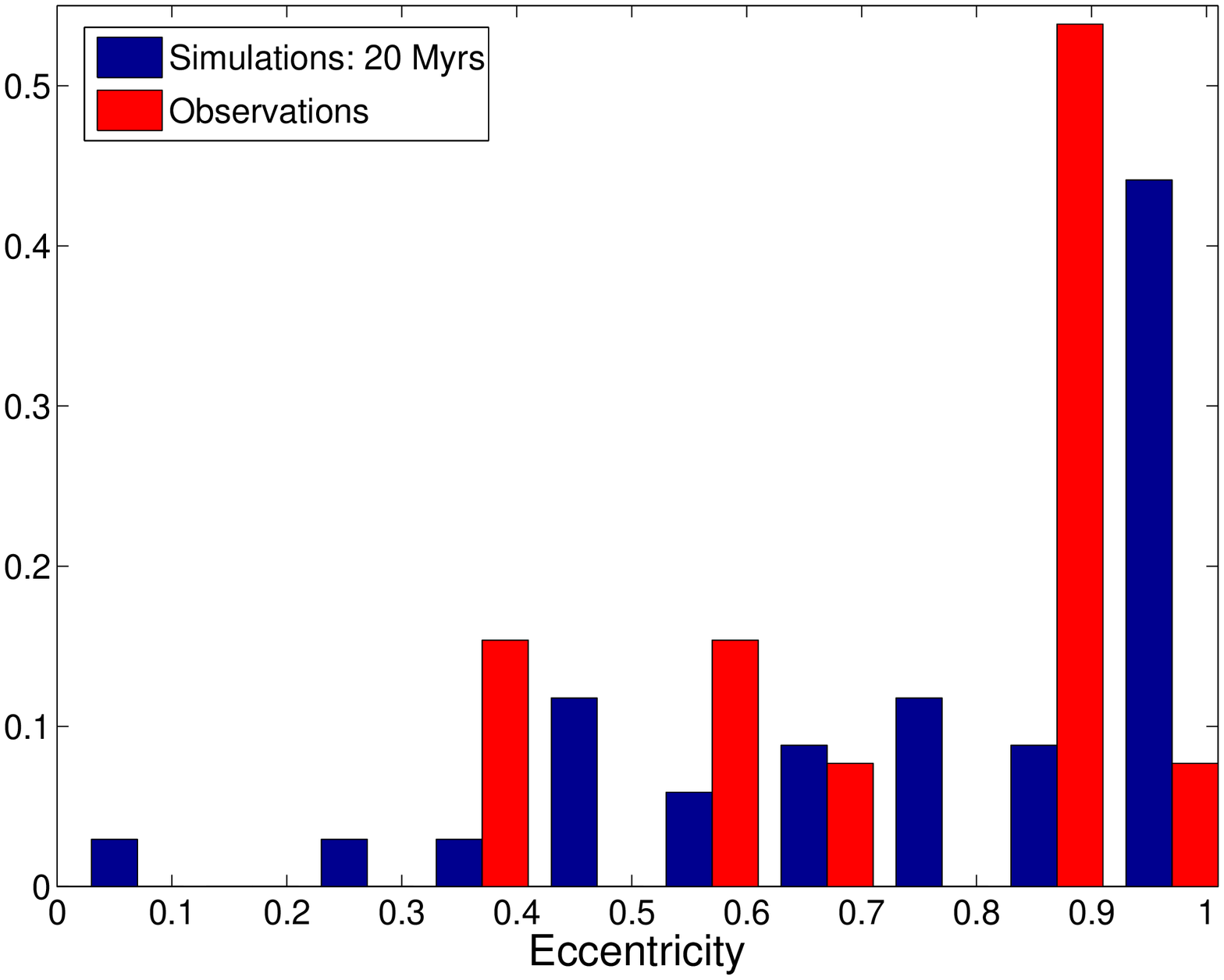}}
\resizebox{\hsize}{!}{\includegraphics[clip=true]{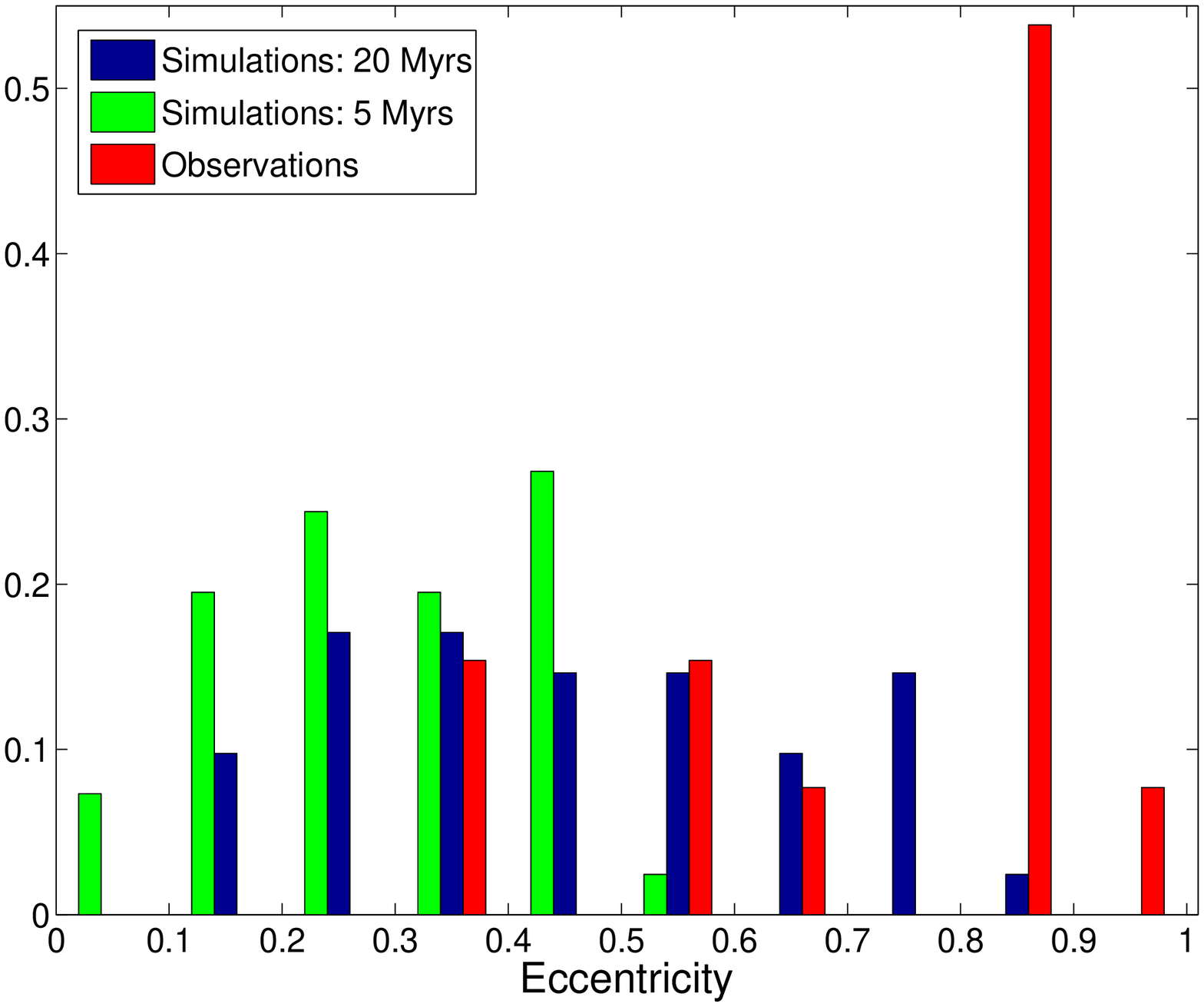}}
\caption{\footnotesize The eccentricities of observed and simulated S-stars
from the binary disruption scenario (after $5$ Myrs; upper figure)
and from the disk migration scenario (after 5 and 20 Myrs; lower figure). }

\label{f1}
\end{figure}

%




\section{The disk stars}

\citet{ale+07} and \citet{cua+08} explored the dynamical evolution
of a single stellar disk using small N-body simulations ($\sim100$
stars), where they studied the effects of massive stars in the disk
(the mass function), and the structure of the disk (eccentric vs.
circular). \citet{cua+08} also studied the role of wide binaries
following \citet{per+08} who suggested binaries could have an important
role in the evolution of the disk, somewhat similar to their role
in stellar clusters \citep{heg75} and in the ejection of OB runaway
stars. These studies showed that although
the different components contribute to the disk evolution, it is difficult
to explain the current eccentricities of the observed stars and the thickness of the disk, with only these components.

We studied a single disk of $\sim5000\, M_{\odot}$, composed of either
$5000$ equal mass stars, or $\sim2500$ stars with a Salpeter mass
function between $0.6-50\, M_{\odot}$ . The initial conditions are
of a thin disk ($H/R\sim0.01)$ with a surface density of $r^{-2}$,
with all stars on circular orbits ($e\lesssim0.01$). In addition
we studied the evolution of such disks both with and without an isotropic
stellar cusp component around the MBH in which the stellar disk is
embedded in. The region of the GC disk is thought to contain a few
$10^{5}$ up to $10^{6}$stars; simulating such a larger number of
stars in orbits close to a MBH, even with a GRAPE cluster is currently
difficult. However, the dynamics and relaxation processes close to
the MBH are dominated mostly by the much smaller number of SBHs in
this region (a few $10^{3}$ up to $10^{4}$ SBHs are though to exist
in the GC disk region ;\citealp{mir+00,hop+06b,fre+06}). Simulating
only this SBHs component could therefore be more efficient in running
time and at the same time capture most of the important relaxation
processes effecting the the dynamics of the disk. In our simulations
we put $1.6\times10^{4}$ SBHs (of $10\, M_{\odot}$ each) with an isotropic
power law distribution $(n\propto r^{-2}$), between $0.01-0.8$ pc. 

The evolution of the mean eccentricity of the disk stars is shown
in Fig. (2), both for low mass stars ($<15\, M_{\odot}$) and high
mass stars ($M\ge15\, M_{\odot}$, i.e. such as the observed disk
stars in the GC). The evolution of a stellar disk embedded in a cusp
of SBHs, is compared with that of an isolated stellar disk with a
Salpeter mass function between $0.6-80\, M_{\odot}$ (i.e. higher
mass cutoff than used in the disk+cusp simulation, to allow for the
disk heating by more massive stars, as discussed in \citealp{ale+07}).
The results show that the SBHs cusp component has an important contribution
to the disk evolution. Although one disk has a lower mass cutoff than
the other, it is heated much more rapidly, due to the contribution
of the cusp stars. More importantly, the high mass stars corresponding
to the stars that are actually observed in the GC disk, have relatively
low eccentricities in the isolated disk, and would present difficulty
to our understanding of the disk evolution, as discussed by \citealp{ale+07,cua+08}.
Adding the cusp component solves the problem, as even the eccentricities
of the higher mass stars are high in this case and comparable with
the observed eccentricities of the disk stars in the GC. We note that
these simulations did not take into account the contribution of low
mass disk stars (i.e. not the SBHs in the cusp) which will further
accelerate the heating of the stellar disk. 

\begin{figure}[h!]
\resizebox{\hsize}{!}{\includegraphics[clip=true]{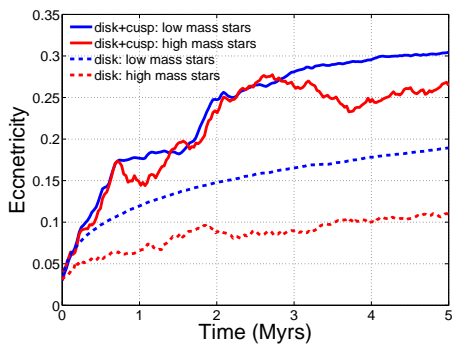}}

\caption{\footnotesize The evolution of the mean eccentricity of the disk stars (both low mass, $M<15\, M_{\odot}$ and high mass stars, $M>15\, M_{\odot}$) with and without interactions with the stellar cusp around the MBH. }
\label{f2}
\end{figure}

\section{Summary}

\label{sec:summary}

The dynamical evolution of the young stars in the GC both in the stellar
disk(s) and in the inner S-cluster is not yet understood. We used
N-body simulations to study the dynamics and origin of these stars.
We found that the S-stars close to the MBH in the GC could be stars
that were captured following a binary disruption by the MBH, and later
on dynamically evolved due to scattering by other stars, or stellar
black holes, to obtain their currently observed orbits. We also show
the the young stellar disk could have formed as a cold (thin) circular
disk and evolve to its currently observed thick (hot) disk, mostly
due to scattering by cusp stars, whereas self relaxation of the disk
plays a more minor role, especially in regard to the more massive
stars seen in observations.

\bibliographystyle{apj}

\end{document}